\documentclass[12pt,conference]{IEEEtran}
\usepackage{graphicx}
\graphicspath{{./images/}}
\usepackage{float} %positions of tables and figures
\usepackage{subfig}
\usepackage{amsmath} %math
\usepackage{amssymb}
\usepackage{url}
\usepackage{hyperref} 
\hypersetup{breaklinks} 
\usepackage{listings}
\usepackage{color}
\usepackage{todonotes}

\renewcommand\cap[3]{\caption[#2]{\label{#1}\textsc{#2}. \small\textit{#3}}}

\long\def\comment#1{}

\begin{document}
%paper title
%AOSPIR
\title{Open Set Intrusion Recognition for Fine-Grained Attack Categorization}

\vspace{-2em}
\author{\IEEEauthorblockN{\small
Steve Cruz\IEEEauthorrefmark{1},
Cora Coleman\IEEEauthorrefmark{2},
Ethan M. Rudd\IEEEauthorrefmark{1}
and Terrance E. Boult\IEEEauthorrefmark{1}}

\IEEEauthorblockA{\small
  \IEEEauthorrefmark{1} University of Colorado Colorado Springs Vision and Security Technology (VAST) Lab \IEEEauthorrefmark{2} New College of Florida\\
  Email: \IEEEauthorrefmark{1} \{scruz,tboult,erudd\}@vast.uccs.edu \IEEEauthorrefmark{2} \{cora.coleman14\}@ncf.edu\\
  Address: 1420 Austin Bluffs Pkwy, Department of Computer Science, Colorado Springs, CO, 80918\\
  Phone: (719) 255-3544 Fax: (719) 255-3369
}
}
\vspace{-2em}

%make the title area
\maketitle

%abstract
\begin{abstract}
  Confidently distinguishing a malicious intrusion over a network is an important challenge. Most intrusion detection system evaluations have been performed in a closed set protocol in which only classes seen during training are considered during classification. Thus far, there has been no realistic application in which novel types of behaviors unseen at training -- unknown classes as it were -- must be recognized for manual categorization. This paper comparatively evaluates malware classification using both closed set and open set protocols for intrusion recognition on the KDDCUP'99 dataset. In contrast to much of the previous work, we employ a fine-grained recognition protocol, in which the dataset is loosely open set -- i.e., recognizing individual intrusion types -- e.g., ``sendmail'', ``snmp\_guess'', ..., etc., rather than more general attack categories (e.g., ``DoS'',``Probe'',``R2L'',``U2R'',``Normal'').
We also employ two different classifier types -- Gaussian RBF kernel SVMs, which are not theoretically guaranteed to bound open space risk, and W-SVMs, which are theoretically guaranteed to bound open space risk.
We find that the W-SVM offers superior performance under the open set regime, particularly as the cost of misclassifying unknown classes at query time (i.e., classes not present in the training set) increases.
Results of performance tradeoff with respect to cost of unknown as well as discussion of the ramifications of these findings in an operational setting are presented.
\end{abstract}

%keywords
\comment{
}

\IEEEpeerreviewmaketitle

\section{Introduction}

Intrusion detection systems aim to ascertain if a system is under attack by either 1.) detecting anomalies or 2.) recognizing signatures of known attack patterns. 
Anomaly detection approaches label behavior that deviates from a ``normal'' model as anomalous, often making an implicit (and flawed) assumption that anomalies correspond to intrusive or problematic events ~\cite{modi2013survey}. However, anomalies can naturally spring up from benign activity, e.g., system re-configurations.
The conflation between \emph{anomalous events} and \emph{intrusive events} often accounts for higher false positive rates of anomaly-based systems. The high false positive rates and lack of diagnostic information are two reasons why anomaly-based systems are seldom deployed in practice~\cite{sommer2010outside}.
Nonetheless, anomaly based approaches offer an important capability that discriminative signature based approaches generally do not: namely, they generalize to novel classes of attacks.

\begin{figure}[!htbp]
  \centering
  \subfloat[\label{fig:openset1}]{\includegraphics[width=0.16\textwidth]{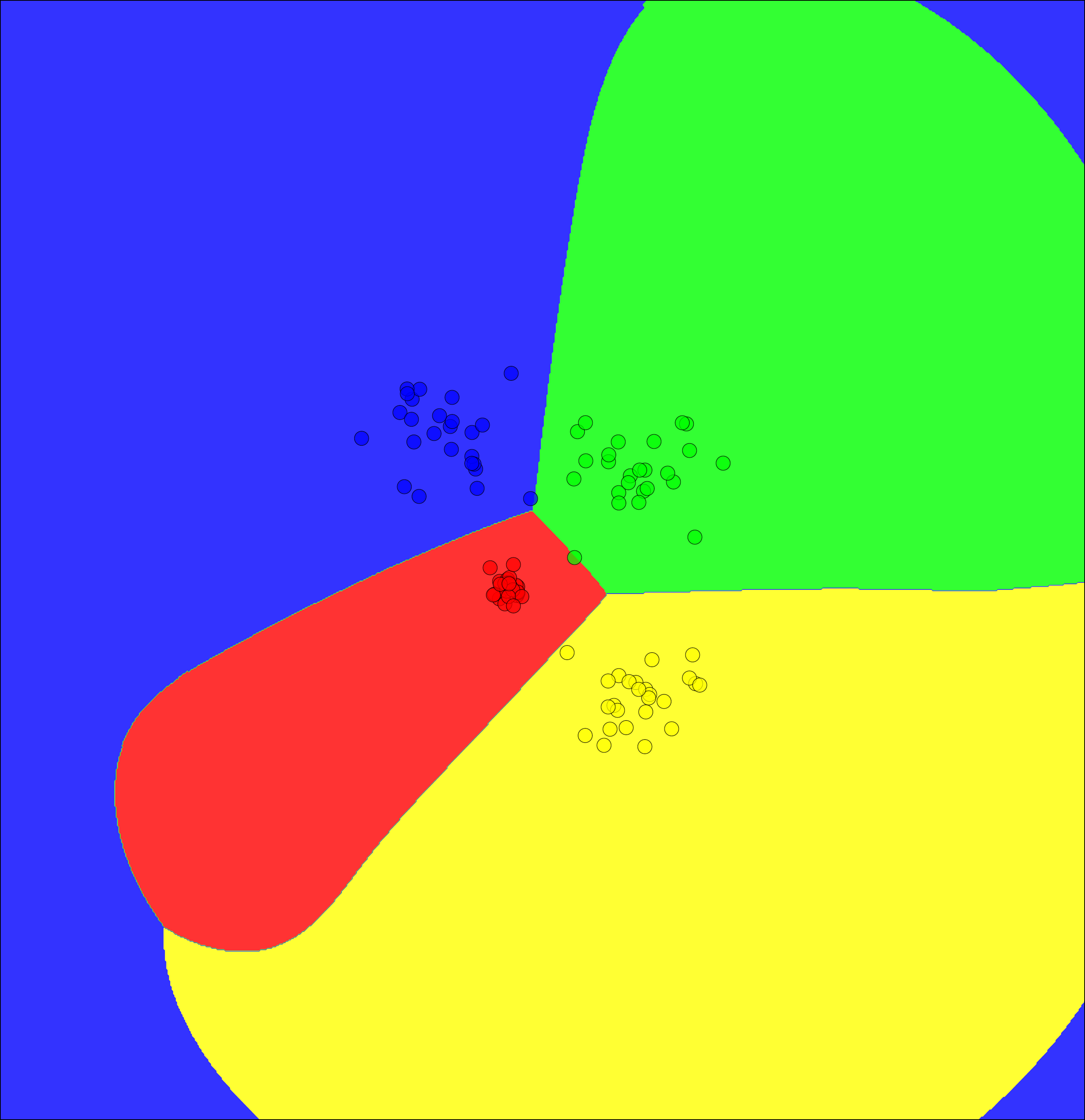}}%\hspace*{.01\textwidth}
  \subfloat[\label{fig:openset2}]{\includegraphics[width=0.16\textwidth]{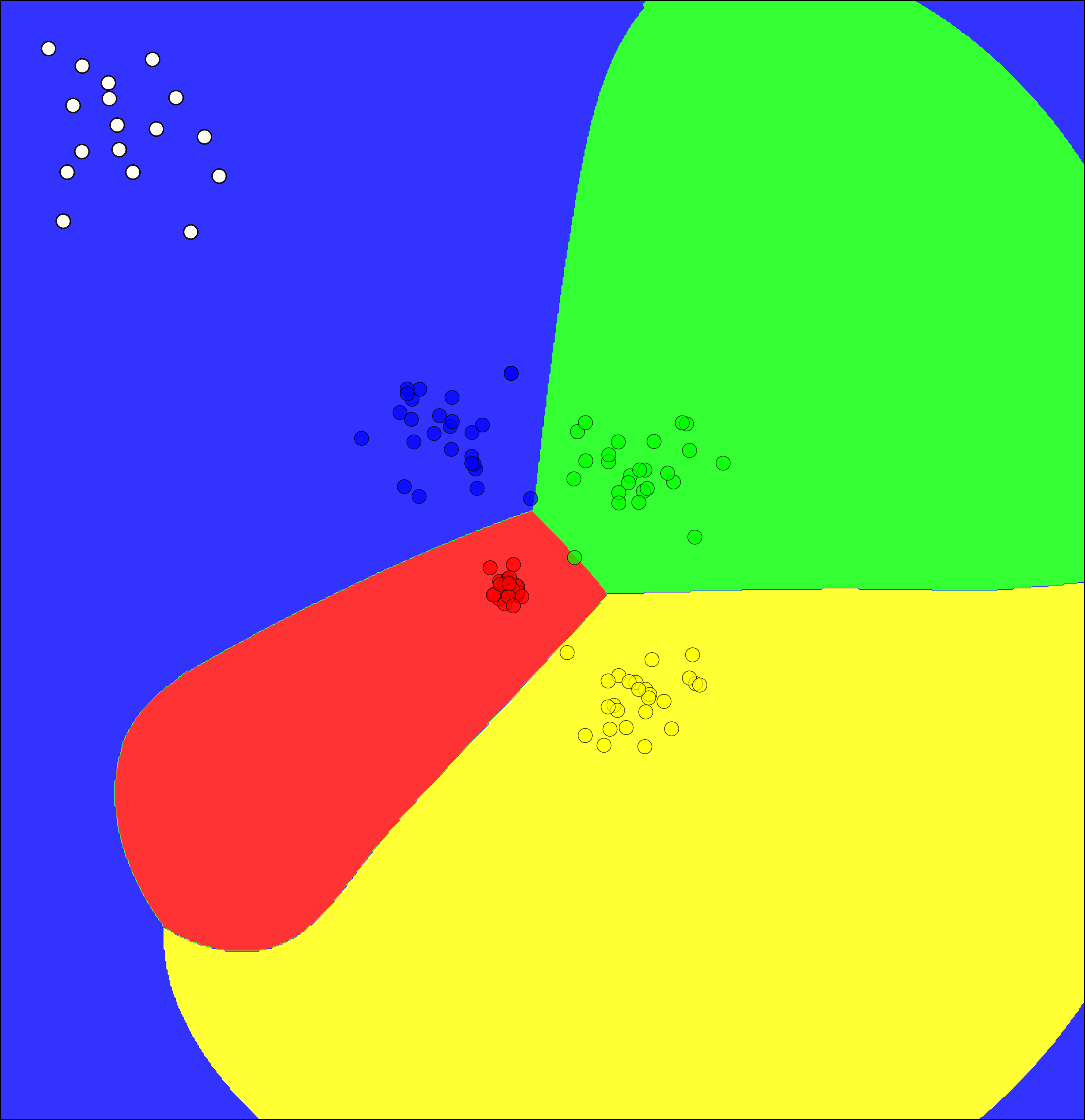}}%\hspace*{.01\textwidth}
  \subfloat[\label{fig:openset3}]{\includegraphics[width=0.16\textwidth]{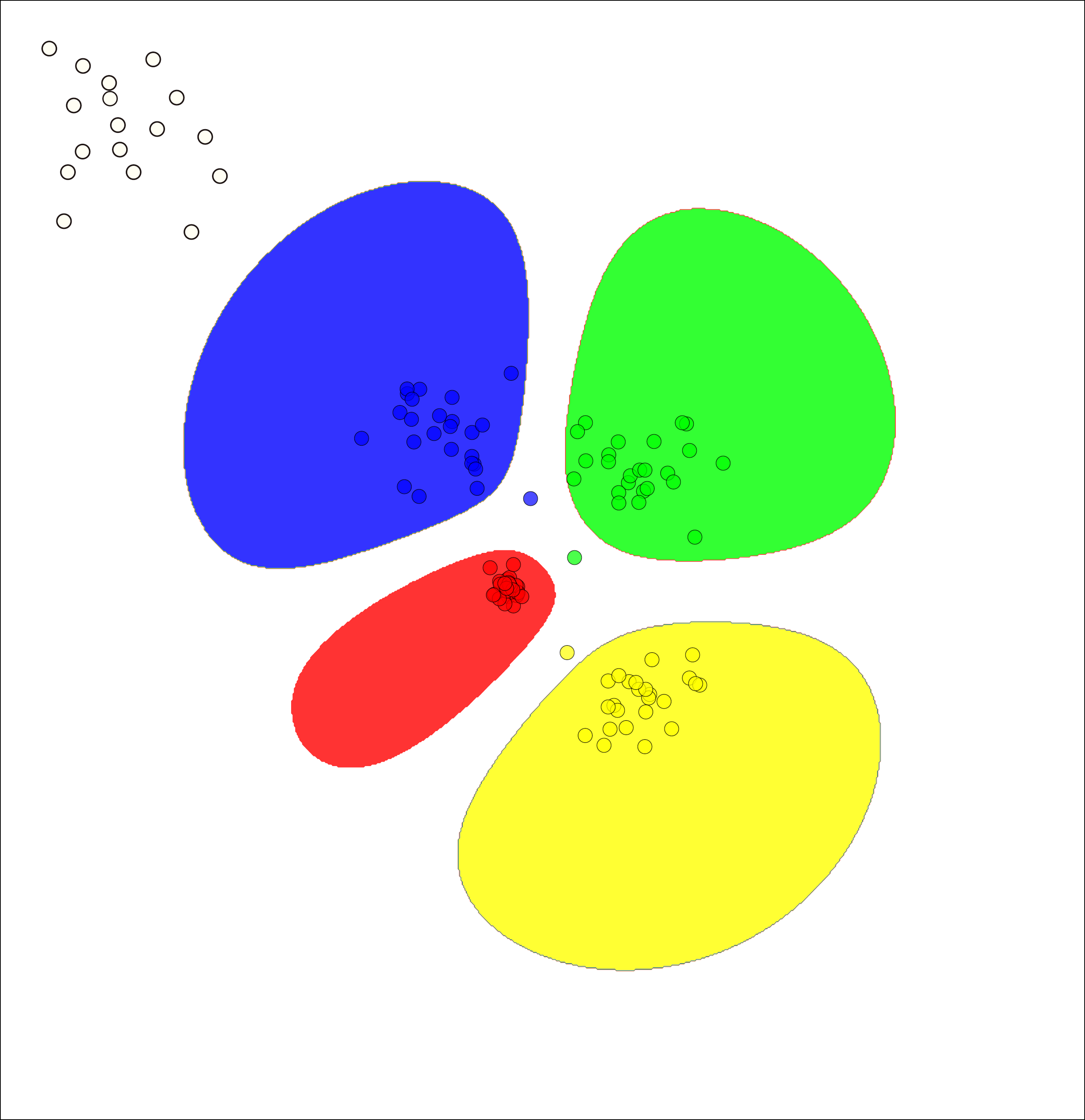}}
  \cap{fig:openset}{The Open Set Problem}{ In \protect\subref*{fig:openset1}, a closed-set classifier is trained to discriminate between four classes (blue, green, yellow, and red). Samples from a different class not in the training set (white) will, at classification time, \protect\subref*{fig:openset2}, be misclassified as blue (and with high confidence if distance-based probability calibration is applied). Open set recognition, \protect\subref*{fig:openset3}, aims to discriminate samples from classes within the closed set, while limiting the scope of its decision by the support of the training data. Via this support bound, samples from a novel class are labeled ``unknown'' rather than mis-ascribed to the wrong class.}
\end{figure}

Most signature methods, by contrast, classify attack patterns based on features from previously seen attack and normal system states.
Several techniques can be employed to perform signature matching, including exact/partial matching, formal grammars, and machine learning~\cite{rudd2016survey}.
One advantage of signature matching is that it can provide diagnostic information by performing not only intrusion detection -- which answers the question ``is an attack occurring?'' -- but also intrusion recognition -- which answers the question ``what kind of attack is occurring?''. 
Of course, the degree of diagnostic information provided depends on the granularity at which the system was trained: training with unspecific labels is insufficient to provide useful diagnostics; rather, a fine-grained multi-class discriminative regime is ideal, in which classification is performed on \emph{individual exploits} instead of on nebulous attack meta-labels. 

Unfortunately, fine-grained labeling requires considerable overhead, and attack databases must be continuously updated for discriminative signature-based methods to work well. 
Despite their other advantages, signature methods do not generalize well to novel attack types. 
As more attack types emerge, one must choose carefully, especially under limited resources, which samples to label and use to  update the classifier. 
Notably, one must make this decision under an \emph{Open Set} assumption -- that samples seen at classification time do not necessarily come from classes seen in the training set.
Thus, both approaches -- anomaly based and signature based -- have their own unique merits and flaws. In this paper we propose a theoretical framework to best reconcile the merits of both systems into one unified framework, which allows fine-grained discrimination between known malware samples and also offers the capability to recognize when samples at classification time come from a previously unseen class.

\section{An Open Set Approach}

To arrive at this unified framework, we borrow from ~\cite{Scheirer:2014,Scheirer:2013}, the concept of \emph{Open Set Recognition}, which aims to balance the minimization of \emph{empirical risk} -- the risk of misclassifying known signatures --  with \emph{open space risk} -- the risk of matching signatures from novel exploits to types of either intrusive or normal behavior not present in the (closed) training set. In the most basic sense, classes utilized in testing are not present in training. Recognition gives the assumption that classes are recognized in a much larger space of classes that are not recognized. When dealing with the open set recognition problem, the data will have multiple known classes and many unknown classes. The unknowns are what make the problem \emph{open set}.  

Under an open set framework, anomaly detection methodologies are employed not to label data as corresponding to a known class (e.g., malicious), but rather to mark query samples that lack statistical support from the training set as ``unknown''. For samples in regions of high support, the decision of the discriminative signature based classifier is used; in regions of low support the sample is marked ``unknown'' for subsequent analysis and potentially retraining/updating the classifier. The ``unknown'' sample is in a region far away from known data, meaning that no other class is nearby to support classification. The Weibull-calibrated SVM (W-SVM) classifier~\cite{Scheirer:2014}, which we use for our experiments, uses statistical extreme value theory (EVT)~\cite{Scheirer:2011}, to provide a probability of sample inclusion within a class. The W-SVM attempts to quantify the region and produce a more natural way of stating how far is far, without prior decision of some ad-hoc threshold.

The distinction between ``known'' and ``unknown'' regions of signature space is particularly important if we wish to leverage the powerful generalization capabilities of machine learning. Discriminative classifiers parameterize decision boundaries in high-dimensional signature space by partitioning the space according to an empirical risk minimization criterion. This often leads to decision regions of unbounded support. Consider the decision boundaries in Fig. ~\ref{fig:openset2}, which ascribe class labels from the closed set of training instances to unlimited space. Properly, a classifier should ``make decisions'' consistent with Fig. ~\ref{fig:openset3}, only labeling instances as originating from a known class if there is data-driven support for these decisions.

\begin{figure*}[!htbp]
  \centering
  \includegraphics[width=1\linewidth]{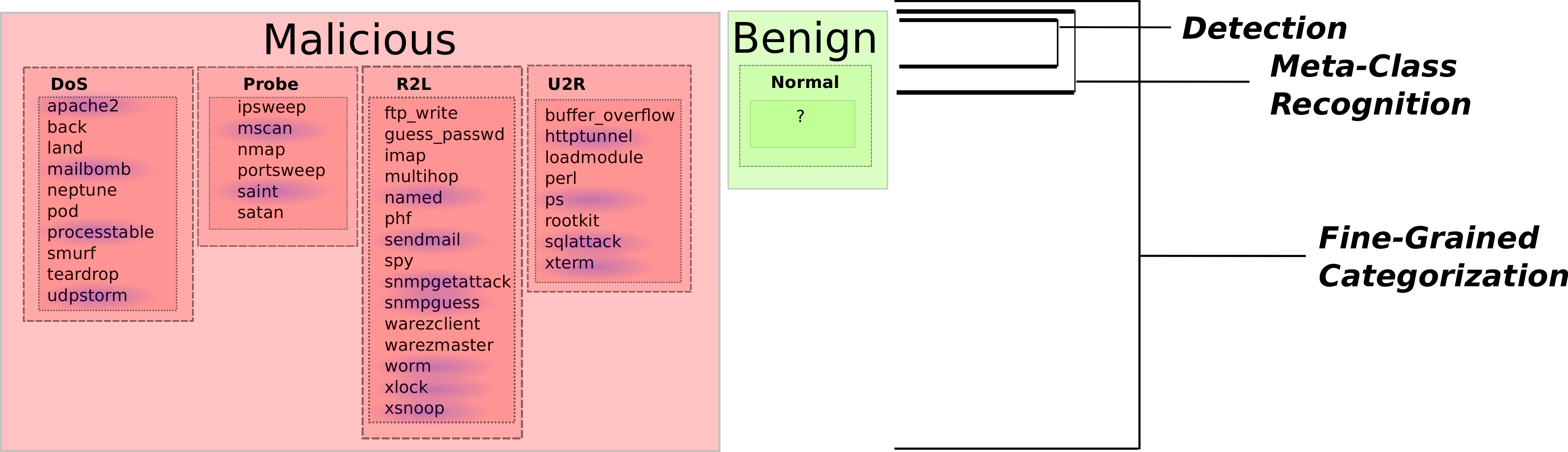}
  \cap{fig:attack_metatypes}{Intrusion Heirarchies}{ 
    Samples can be labeled at many granularities, e.g., fine-grained categorization of individual exploits, attack metatypes, and detection level, i.e., malicious/benign. For the KDDCUP'99 dataset, shown in this figure, exploits highlighted in blue are only present in the test set, not the training set. Because of this discrepancy between training and test sets, an open set protocol is necessary when performing classification at the exploit level. Although it was not their intent, the designers of the KDDCUP'99 naturally ended up in an open set scenario where certain types of attacks only occurred at query time.}
\end{figure*}

Note also that the intrusion detection problem is really a generalization of the intrusion recognition problem. While intrusion detection may appear closed-set insofar as there are only two classes, i.e., intrusive and benign, by giving up fine-grained categorizations for ease of implementation, the problem of novel samples arriving at unsupported regions of feature space still holds. 
Thus, approaches that simplify the intrusion recognition problem to a detection problem for ease of closed-set treatment do not actually address the open set problem; they ignore it.
Open set exists even in a closed-set problem with just malicious or benign samples. Moreover, we believe that instead of over-simplifying the problem, we will explore a more meaningful level of fine-grained recognition, since it yields important diagnostic information about attack class types. The problem in the world today is not just that an attack exists, but what kind of attack and what can be done. Something to take into account as well are the findings from ~\cite{rudd2016survey}. Rudd et al. surveyed malicious stealth technologies and existing autonomous countermeasures. Their findings suggest that while machine learning has potential for generic and autonomous solutions, several flawed assumptions fundamental to most recognition algorithms inhibit a discrete mapping between the stealth malware recognition problem and a machine learning solution. The closed set assumption was the most notable of these flawed assumptions. Unseen classes at classification time exist for real-world intrusion recognition tasks, and neither all variations of malicious code nor all variations of harmless behavior can be known apriori. 

\section{Experiments} 

To demonstrate the gains obtained from moving to an open set regime, we performed an open set analysis on the KDDCUP'99 dataset -- one of the most popular datasets in intrusion detection literature ~\cite{Tsai:2009}.
We selected this dataset because it is widely used, and offers a basis for comparison of our framework. 
However, this dataset is old and does not accurately reflect a modern network environment ~\cite{sommer2010outside}, so our experiments should be viewed as a verification of the theory behind our framework, not as an operational assessment of a deployable system. 
No research that we could find uses the KDDCUP'99 dataset in an exploit-level open set evaluation. 
We suspect that this is because, when working at the detection level, or the meta-class recognition level, the problem can be over-simplified to closed-set, which assumes that any instance showing up at query time will lie in a region of ``known'' support.
However, as we see in Fig.~\ref{fig:attack_metatypes}, when operating at the individual exploit level, there are a number of distinct exploit types in the test set that do not occur in the training set. Other algorithms have had a difficult time doing well with the dataset because it is still not a solved problem. 
Thus when working at the finer exploit-level granularity, we cannot ignore the open set problem and \emph{should} operate under an open set protocol. 

The dataset consists of connection vectors of 41 features, each of which is labeled according to Fig. ~\ref{fig:attack_metatypes}. Prior to conducting our evaluation, we followed standard practice of removing the large number of redundant records in the dataset~\cite{Portnoy:2001,Tavallaee:2009}. Tavallaee et al. analyzed the KDDCUP'99 dataset and discovered that 78\% and 75\% of the records are duplicated in the train and test set, respectively. Within the train set, these redundancies will cause learning algorithms to be biased towards more frequent records, which in turn hinder it from learning infrequent samples. Infrequent samples are typically more harmful to networks, for example, User to Root Attacks (U2R). Within the test set, redundancies will cause the evaluation results to be biased towards methods that have better detection rates on the more frequent records. The reduction reduced the training set from 4898431 entries to 1074974 entries and reduced the test set from 311029 entries to 77216 entries. We noticed that two of the classes within the training set still accounted for a disproportionate amount of the data, so we downsampled these classes by a factor of 100. Classes with fewer than 20 samples we removed entirely from the training set. To format the data for learning we normalized each vector element to a 0-1 range using a linear min-max scaling across all data on an element-wise basis. For the categorical data, we assigned integer values corresponding to an index in a number of categories prior to conducting the scaling.
 
\begin{figure}
 \centering
  \includegraphics[width=\linewidth]{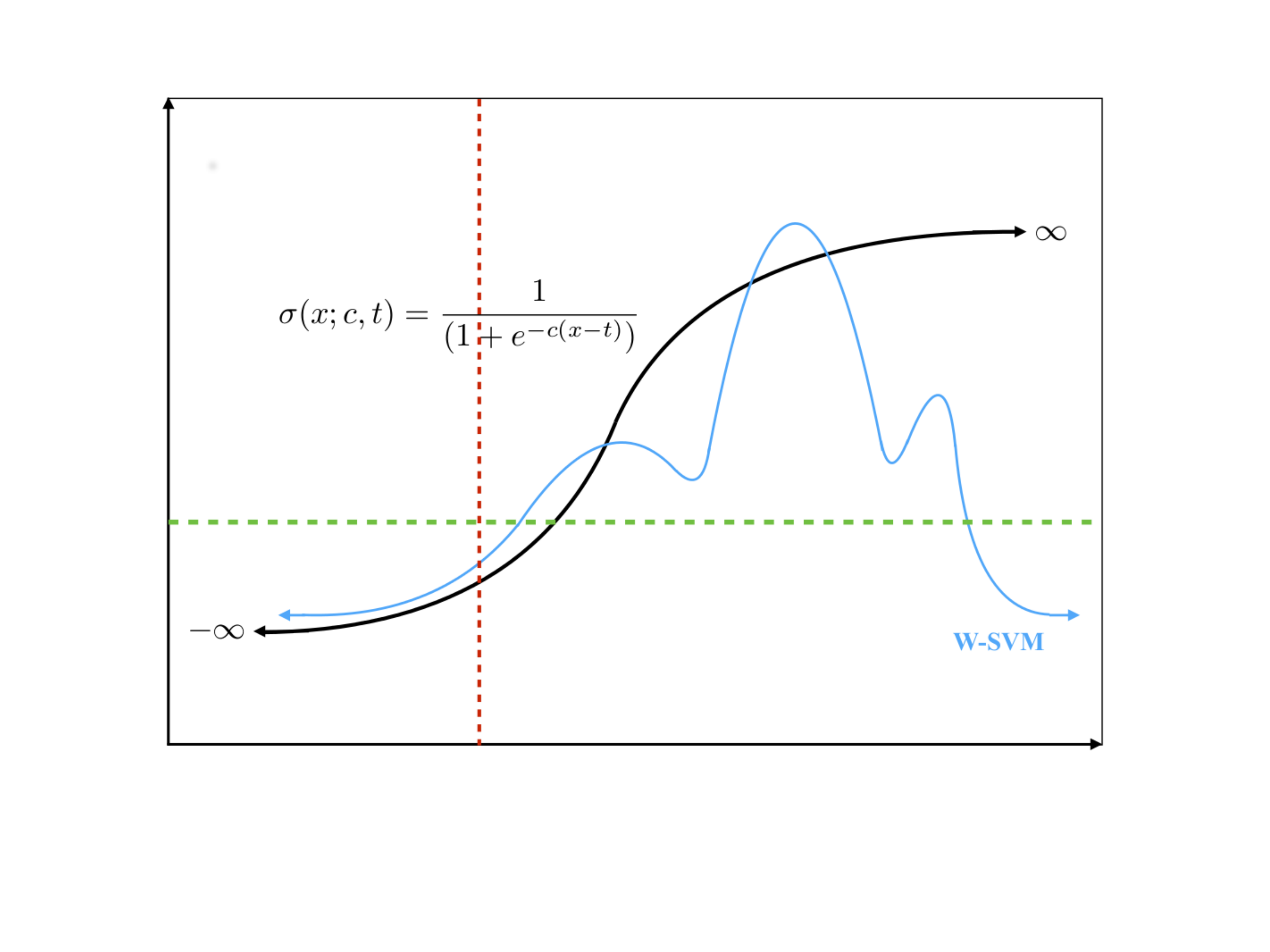}
 \caption{\small Let red correspond to the decision boundary. Platt calibration (sigmoidal) with respect to distance from the decision boundary of an RBF SVM does not bound open space risk because when thresholded (green), the calibration still labels unbounded space with approximately 100\% probability. Probabilities of inclusion returned from a Weibull-calibrated W-SVM (e.g., blue) abate to zero with distance from the actual data, and thus provably bound open space risk.}
 \label{fig:calibrations}
\end{figure}

In our analysis, we performed a comparison between two analogous classifiers: a multi-class closed set Gaussian RBF kernel Platt-calibrated~\cite{Platt:1999} SVM, and a multi-class  open set Gaussian RBF kernel W-SVM~\cite{Scheirer:2014}. For the Platt-calibrated SVM, a sigmoid is used to normalize scores to reflect probabilities (i.e., scores are monotonically increasing with distance from the decision boundary), whereas the W-SVM's \textit{compact abating probabilities} (CAP models) are mathematically guaranteed to abate away from regions of known support (implementation details can be found in ~\cite{Scheirer:2014}). This is also shown in Fig.~\ref{fig:calibrations}.

In order to determine a good value for $C$ and $\gamma$ in our experiments, we first performed a 3-fold cross validated order of magnitude grid search over the training set. We used a grid of $C,\gamma \in \{10^-5, 10^-3, \hdots, 10^5\}$. Based upon accuracy over this grid search, we selected a $C$ value of 1000 and a $\gamma$ value of 0.1 for training our classifiers. 

We evaluated classification error in terms of both closed and open set accuracy. For the closed set protocol this is simply the number of correctly classified test instances out of the total number of test instances, which does not account for unknown classes: i.e., all test instances with labels not present in the training set will be misclassified under this protocol. For the Platt-calibrated RBF SVM we obtained a closed set accuracy of 91.1\%. For the W-SVM we obtained a closed set accuracy of 90.1\%. A common misconception is that the start off accuracy difference should theoretically yield the same closed-set accuracy for both classifiers, however, that is not the case because each classifier utilizes a different model. Also, we attribute the discrepancy to implementation-level differences introduced by Scheirer et al. in the W-SVM solver~\cite{Scheirer:2014}.

For evaluating under an open set protocol we thresholded probabilities returned by both W-SVM and the Platt-calibration on the RBF SVM by a variety of different thresholds. We do not address threshold selection in this paper, although this has been discussed in several other works including~\cite{Scheirer:2013} and~\cite{Scheirer:2014}. Instead, we inspected results under a variety of thresholds from 0.1 to 0.3. We found that both of these thresholds close the accuracy gap such that each classifier performs at 91.5\% open set recognition accuracy, indicating that the open set W-SVM sees gains over the closed set Platt calibrated SVM when evaluated under an open set protocol. However, these gains are limited because despite the \textit{number of classes} in the test set not seen in the training set (cf. Fig. ~\ref{fig:attack_metatypes}), over 95\% of classes are still from known samples. To indicate the \textit{actual} gains of the open set classifier for unknown classes, we introduce the concept of \textit{``Cost of Unknown''} -- i.e., the impact on perceived accuracy with respect to number of unknown samples. This is achieved by weighting the relative importance of accuracy on known and unknown partitions of the test data -- e.g., meaning a 10\% cost of unknown weight open set accuracy on unknown classes in the test set at 0.1 and closed set accuracy of known classes at 0.9. Fig.~\ref{fig:results}  depicts perceived accuracy as the cost of unknown increases. When accuracy on unknown samples is given no weight, the Platt-calibrated SVM offers slightly better performance, but as we sligtly upweight the cost of unknown to just 7\%, (i.e., accurate classification of unknown is given 7\% weight, whereas accurate classification of known is given 93\%) the W-SVM's performance quickly eclipses that of the RBF SVM, with an increasing performance gap as cost of unknown increases. 

\begin{figure}[h!]
 \centering
  \includegraphics[width=0.552\textwidth]{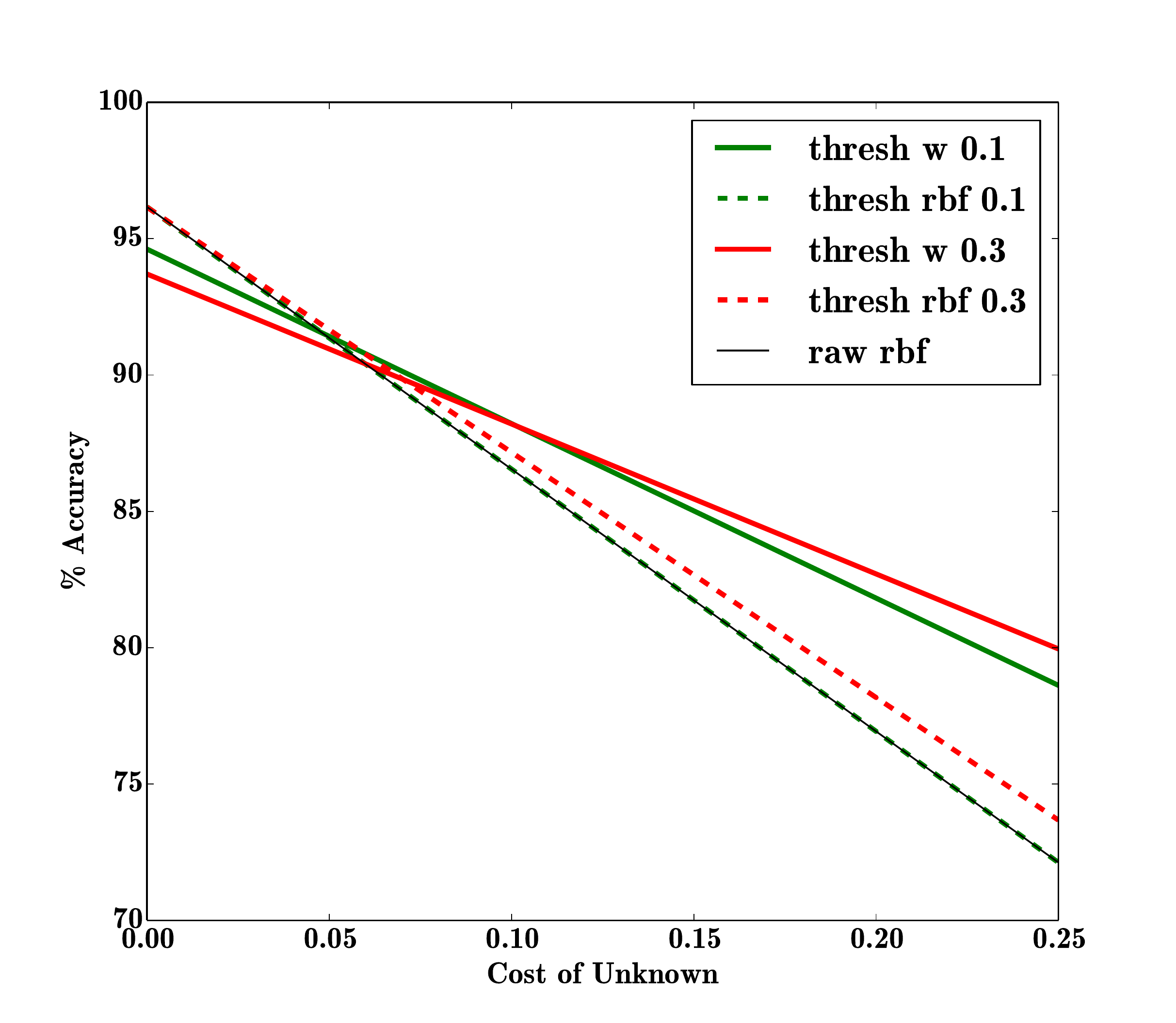}
  \cap{fig:results}{Open Set Evaluation}{ 
    This figure depicts decay in accuracy as we up-weight the cost of labeling an unknown instance as coming from a known class for both open-set and closed-set classifiers. Labels \emph{thresh RBF} and \emph{thresh W} correspond to Platt-calibrated and W-SVM classifiers respectively. As we can see, the open set W-SVM does a better job of rejecting unknown samples for both 0.1 and 0.3 probability thresholds.
}
\end{figure}

\section{Discussion}

Our evaluation demonstrates that open set recognition has potential to benefit intrusion recognition applications. Curves like the one in Fig.~\ref{fig:results} can serve to choose between open-set and closed-set classifiers for the application at hand. The closed set classifier performed comparably to the open set classifier on the KDDCUP'99 dataset open set protocol, when considering both known and unknown classes, closing the performance gap in the purely closed set regime. However, it is important to consider that the majority of the KDDCUP'99 test partition is dominated with ``known'' benign samples, which may well be anomalous with respect to the support of the training data. With that in mind, amidst a sea of non-malicious traffic are anomalies that can compromise the secure state of the network. Thus, the \textit{cost} of mislabeling an unknown sample is highly dependent on operational constraints. Once a sample has been classified as unknown, the instance can be blocked until further review. A security expert should then review the sample, classify it accordingly, and train a newer version of the classifier. In the future, whether that sample is malicious or benign, the classifier should detect anything similar to its type. In a perfect world, the system would not need any human input, but current methods cannot achieve that. Implementing this into an existing intrusion recognition system would be ideal as our experiments offer the first experimental proof points that suggest using an open-set approach has some potential for distinguishing novel types of behavior. 

An important topic for future work is assessing the quantity of unknown samples that one would expect to see in realistic environments. 
This number can vary considerably depending on the scenario, and quantifying ``unknown'' can be difficult when different classes of data are ill-defined -- e.g., \textit{should} a previously unseen form of novel ``normal'' behavior be labeled as unknown? According to our open set protocol the answer is no, because we only have labels of finite granularity, but in a realistic operational setting, the story may be different. Other factors such as environment and security levels might also come into play.

Finally, we mention that KDDCUP'99 is a very old dataset and offers little to no utility as a benchmark for the performance of intrusion detection systems on a modern network. At the time of publication, we were unaware of any newer and related datasets. We used it in this paper merely to demonstrate the gains attained by incorporating open set recognition protocols at fine granularity on a canonical network intrusion detection dataset. A more realistic dataset would allow for more thorough evaluation of our open set protocol.

The creation of such a dataset by academia is nearly impossible. Privacy policies and concerns are the biggest barrier for the formulation of an updated and improved dataset. Simulating real network traffic with real day-to-day users at a large scale is relatively easy, but unpractical. A university would simply not allow student's confidential data to be utilized by any means.     Future work will require performance assessments on more realistic datasets where gains from an open set recognition system could be significantly higher, and many of these datasets must come from collaboration with industry. The degree to which open set recognition works well is largely a function of the dataset.

Overall, the RBF SVM outperformed the W-SVM on a purely closed set evaluation of the data, but we saw a shift in applying an increasingly open set evaluation where the W-SVM took the lead in performance accuracy. In the future, these techniques should be extended to a better dataset and re-evaluated thereon. We demonstrated that when the cost of unknown increases, however, by applying an open set protocol, we were able to garner useful results that closed set classifiers cannot deliver.

\section*{Acknowledgements}
This work support was supported in part by NSF grant IIS-1320956, the Research Experiences for Undergraduates (REU) program at the University of Colorado Colorado Springs (NSF Award No. 1359275), and the University of Colorado Colorado Springs Undergraduate Research Academy (URA).

\bibliographystyle{IEEEtran}
\bibliography{sources}

\end{document}